\documentclass[preprint,aps,prd,showpacs]{revtex4}
\usepackage{graphicx}
\begin{document}
\draft
\title{Search for electromagnetic properties of the neutrino in $\gamma e$ and $\gamma\gamma$ collisions at CLIC }
\author{A. Senol}
\email{asenol@kastamonu.edu.tr} \affiliation{Department of Physics,
Kastamonu University, 37100, Kastamonu, Turkey}

\begin{abstract}

We have examined nonstandard $\gamma\nu\bar\nu$ and
$\gamma\gamma\nu\bar\nu$ couplings via $\nu\bar\nu$ production in
$e\gamma$ and $\gamma\gamma$ collisions at the CLIC. We obtain 95 \%
confidence level bounds on $\gamma\nu\bar\nu$ and
$\gamma\gamma\nu\bar\nu$ couplings by considering the backscattered
photon distribution function for incoming photons and Initial State
Radiation (ISR) and Beamstrahlung (BS) effect for initial state
electrons in the $e\gamma$ and $\gamma\gamma$ collider modes of
linear collider. We indicate that the reaction
$\gamma\gamma\to\nu\bar\nu$ provides more than 15 orders of
magnitude improvement in neutrino-two photon couplings compared to
LEP limits.
\end{abstract}
 \maketitle
\section{INTRODUCTION}
Searches on electromagnetic properties of neutrinos have became one
of the important issue in particle physics, astrophysics and
cosmology after observation of neutrino oscillations gives evidence
that the neutrino mass is non-zero
\cite{Fukuda:1998mi,Ahmad:2002jz,Eguchi:2002dm}. Moreover, this
searching may play a key role to understanding the physics beyond
the Standard Model (SM) and contributes to studies in astrophysics
and cosmology. Although, the coupling of neutrino-photon ($\nu\bar
\nu \gamma$) and neutrino two-photon ($\nu\bar \nu\gamma\gamma$)
interactions by evaluating radiative diagrams in the simplest
extension of SM with massive neutrinos are very small, there are
several models beyond SM, predicting relatively large coupling.
Thus, one of the best method to research electromagnetic properties
of neutrinos is a model-independent way.

Anomalous properties of neutrinos will already be well known from
The CERN Large Hadron Collider (LHC) data before a TeV scale linear
collider runs, since the LHC can produce very massive new particles
and will extend the possibilities of testing for new physics
effects. Nevertheless, the LHC may not provide precision
measurements as a result of the typical characteristic of hadron
machine. Whereas, a linear collider with energies on the TeV scale,
extremely high luminosity and clean experimental environment, can
provide complementary information for these properties with
performing precision measurements that would complete the LHC
results. A most popular proposed linear colliders with energies on
the TeV scale ($\sqrt{s}=$ 3 TeV) and extremely high luminosity
($\mathcal{L}=5.9\cdot10^{34}$ cm$^{-2}$s$^{-1}$) is Compact Linear
Collider (CLIC) \cite{Braun:2008zzb}. CLIC generates an accelerating
gradient of 150 MVm$^{-1}$ with the resulting 20 km of active
length. CLIC uses two beam accelerator technology operating at 30
GHz radio frequency to reach this high accelerating gradient. In
addition to $e^+e^-$, linear colliders provide a suitable platform
to study $\gamma\gamma$ and $\gamma e$ interactions at energies and
luminosities comparable to those $e^+e^-$ collisions through the
laser backscattering procedure \cite{Ginzburg:1981vm,
Ginzburg:1982yr}.

The most sensitive experimental bounds on neutrino magnetic moment
are obtained from neutrino-electron scattering experiment with
reactor neutrinos
\cite{Li:2002pn,Daraktchieva:2005kn,Wong:2006nx,Wong:2010pb} and
solar neutrinos \cite{Arpesella:2008mt}, where these limits are
about order of $10^{-11}\mu_B$. Another bounds on magnetic moment of
neutrinos derived from energy loss of astrophysical objects give
about an order of magnitude more restrictive constraint than reactor
and solar neutrino probes
\cite{Raffelt:1999gv,Castellani:1993hs,Catelan:1995ba,
Ayala:1998qz,Barbieri:1988nh,Lattimer:1988mf,Heger:2008er}.

Although, $\gamma\nu\bar\nu$ coupling attract too much attention,
$\gamma\gamma\nu\bar\nu$ coupling has been much less studied in
literature. Current experimental bound on $\gamma\gamma\nu\bar\nu$
coupling was obtained from the LEP data via $Z\to \nu \bar{\nu}
\gamma\gamma$ decay as follows \cite{Larios:2002gq}:
\begin{eqnarray}
\label{leplimit} \left[\frac{1 GeV}{\Lambda}\right]^6
\sum_{i,j,k}\left(|\alpha^{ij}_{Rk}|^2+|\alpha^{ij}_{Lk}|^2\right)\leq2.85\times10^{-9}
\end{eqnarray}
and the analysis of Primakoff effect on $\nu_\mu N\to \nu_s N$
conversion in the external Coulomb field of the nucleus $N$ yields
about two orders of magnitude more restrictive bound than $Z\to \nu
\bar{\nu} \gamma\gamma$ decay. In refs.
\cite{Sahin:2010zr,Sahin:2012zm}, the potential of $\gamma p$ and
$\gamma\gamma$ collisions at the LHC was studied to probe
neutrino-photon and neutrino-two photon coupling. In ref.
\cite{Sahin:2012zm}, it was shown that the reaction $p p\to p\gamma
p\to p \nu \bar \nu q X$ provides more than eight orders of
magnitude improvement in $\gamma\gamma\nu\bar\nu$ couplings compared
to LEP limits.

The purpose of the present paper is to report on the possibility of
obtaining indirect bounds on electromagnetic properties of neutrinos
from anomalous $\gamma\nu\bar\nu$ and $\gamma\gamma\nu\bar\nu$
couplings via $e^-\gamma\to\nu\bar\nu e^-$ and
$\gamma\gamma\to\nu\bar\nu$ processes at the CLIC.
\section{Effective lagrangian for $\gamma\nu\bar\nu$ and $\gamma\gamma\nu\bar\nu$ interactions}
The magnetic moment couple neutrinos to photon through the
dimension-6 effective Lagrangian term
\cite{Larios:2002gq,Larios:1995zn,Maya:1998ee,Bell:2005kz}
\begin{eqnarray}
\label{eqn1} {\cal
L}=\frac{1}{2}\mu_{ij}\bar{\nu}_{i}\sigma_{\mu\nu}\nu_{j}F^{\mu\nu}
\end{eqnarray}
where, $F^{\mu\nu}$ is the electromagnetic field tensor, $i,j$ are
the flavor indices, $\mu_{ii}$ is the magnetic moment of $\nu_i$ and
$\mu_{ij}$ $(i\neq j)$ is the transition magnetic moment. The new
physics energy scale $\Lambda$ is embedded in the definition of
$\mu_{ij}$ in the above dimension-6 effective Lagrangian.

Now we turn to the neutrino-two-photon ($\gamma\gamma\nu\bar\nu$)
vertex which describing from following dimension-7 effective
Lagrangian
\cite{Larios:2002gq,Dodelson:1991dt,Nieves:1982bq,Ghosh:1982rt,Liu:1991rg,Gninenko:1998nn}:
\begin{eqnarray}
\label{eqn2} {\cal
L}=\frac{1}{4\Lambda^3}\bar{\nu}_{i}\left(\alpha^{ij}_{R1} P_R+
\alpha^{ij}_{L1} P_L\right)\nu_{j}\tilde
{F}_{\mu\nu}F^{\mu\nu}+\frac{1}{4\Lambda^3}\bar{\nu}_{i}\left(\alpha^{ij}_{R2}
P_R+ \alpha^{ij}_{L2} P_L\right)\nu_{j} F_{\mu\nu}F^{\mu\nu}
\end{eqnarray}
where $P_{L(R)}=\frac{1}{2}(1\mp\gamma_5)$, $\tilde
{F}_{\mu\nu}=\frac{1}{2}\epsilon_{\mu\nu\alpha\beta}F^{\alpha\beta}$,
$\alpha^{ij}_{Lk}$ and $\alpha^{ij}_{Rk}$ are dimensionless coupling
constants. We will focus on Dirac neutrino case and obtain model
independent bounds on couplings using effective Lagrangians
(\ref{eqn1}) and (\ref{eqn2}).

During calculations in this paper we set $\Lambda$= 1 GeV as a
reference value. In order to prevent any misunderstanding, we should
note that this does not means new physics appears at 1 GeV which is
a quite low value for a new physics energy scale. We set $\Lambda$=
1 GeV as a reference value which makes our bounds easy to compare
with current experimental bounds (see Eq. (\ref{leplimit})). Bounds
at any value for $\Lambda$ can easily be extracted from our tables
by multiplying an appropriate factor.

\begin{figure}[htbp!]
 \includegraphics[width=17cm]{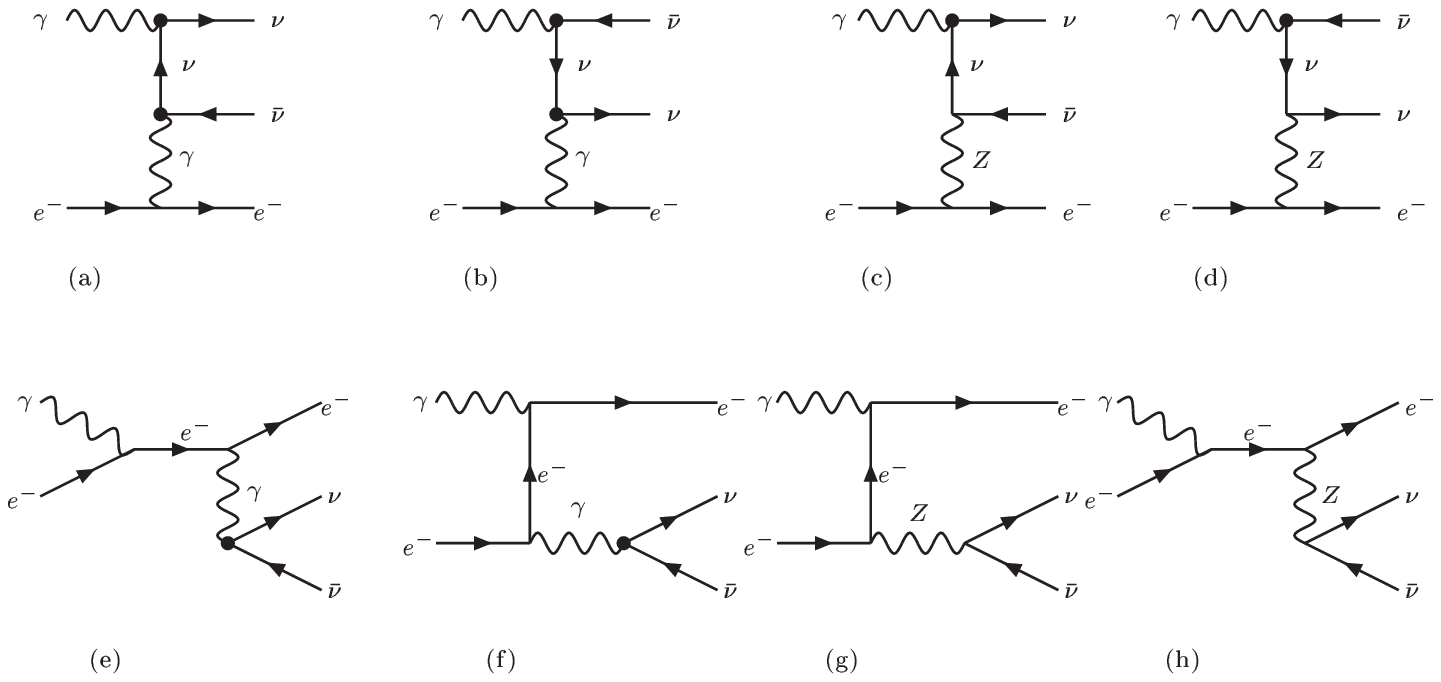}
 \caption{Tree-level Feynman diagrams for the process $e^-\gamma\to\nu\bar\nu e^-$ in the existence of non-standard $\nu\bar\nu\gamma$
 coupling. The dots denote non-standard $\nu\bar\nu\gamma$
 couplings.
 }\label{fig1}
\end{figure}

\begin{figure}[htbp!]
 \includegraphics[width=15cm]{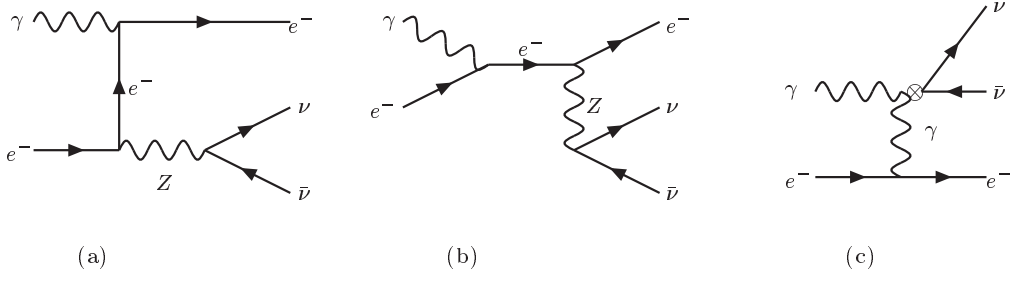}
 \caption{Tree-level Feynman diagrams for the subprocess $e^-\gamma\to\nu\bar\nu e^-$ in the existence of non-standard $\nu \bar \nu \gamma
\gamma$ coupling. The $\otimes$ denote non-standard $\nu \bar \nu
\gamma \gamma$ couplings. }\label{fig2}
\end{figure}

In this study, we consider $e^-\gamma\to\nu\bar\nu e^-$ and
$\gamma\gamma\to\nu\bar\nu$ processes for searching non-standard
$\nu\bar\nu\gamma$ and $\gamma\gamma\nu\bar\nu$ interactions which
denotes Lagrangians (\ref{eqn1}) and (\ref{eqn2}). The complete sets
of Feynman diagrams  contributing to $e^-\gamma\to\nu\bar\nu e^-$ at
tree level are shown in Figs. \ref{fig1} and \ref{fig2}. As seen
from Fig.\ref{fig1}, the first six diagrams((a)-(f)) contains
nonstandard $\nu\bar\nu\gamma$ vertices and the last two diagrams
((g) and (h)) comes from SM electroweak processes.

\begin{figure}[htbp!]
 \includegraphics[width=15cm]{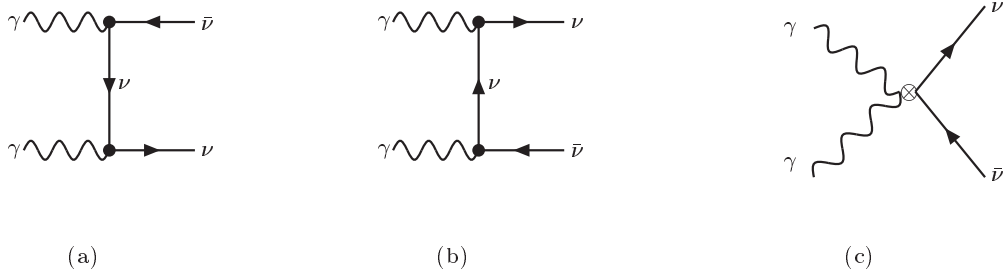}
 \caption{Tree-level Feynman diagrams for the subprocess $\gamma\gamma\to\nu\bar\nu$ in the existence of non-standard both $\nu\bar\nu\gamma$ and $\nu \bar \nu \gamma
\gamma$ couplings. The dots and $\otimes$ denote non-standard
$\nu\bar\nu\gamma$ and $\nu \bar \nu \gamma \gamma$ couplings,
respectively.}\label{fig3}
\end{figure}

As shown in Fig.\ref{fig2}, we have only three Feynman diagrams in
the presence of the effective interaction (\ref{eqn2}). Fig.
\ref{fig3} shows the Feynman diagrams of $\gamma\gamma\to\nu\bar\nu$
process which consists of only non-standard $\nu\bar\nu\gamma$ and
$\gamma\gamma\nu\bar\nu$ interaction vertices.
In order to examine all numerical calculations, we have implemented
the $\gamma\nu\bar\nu$ and $\gamma\gamma\nu\bar\nu$ vertices into
the tree-level event generator CompHEP \cite{Pukhov:1999gg}.
In this study, we consider Initial State Radiation (ISR) for
incoming electron. ISR is a process of photon radiation by the
incoming electron due to its interaction with other collision
particle. In addition, we take into account this spectrum in our
calculations by using the CompHEP program with beamstrahlung
spectra. Beamstrahlung is a process of energy loss by the incoming
electron due to its interaction with the positron (electron) bunch
moving in the opposite direction. Beamstrahlung spectrum, which
depends on the bunch geometry, bunch charge and the collision
energy, is an attribute of the linear collider design. When
calculating these effects we take into account the beam parameters
of the CLIC as shown in Table~\ref{tab:table_bs}.
\begin{table}
\caption{The beam parameters of the two energy options of the CLIC.
N is the number of particles in the electron bunch, $\sigma_{x,y,z}$
are the average sizes of the electron bunches and $\mathcal{L}$ is
the design luminosity.} \label{tab:table_bs}\centering
\begin{tabular}{llllllll}\hline\hline
parameter             & &$\sqrt s$=0.5 TeV     && $\sqrt s$=3 TeV
\\\hline
$\mathcal{L}$(cm$^{-2}$s$^{-1}$)         && $2.3\cdot10^{34}$    &&
 $5.9\cdot10^{34}$ \\
N(10$^{9}$)           && 6.8     && 3.72  \\
$\sigma_{x}($nm$)$     && 200      && 45   \\
$\sigma_{y}($nm$)$     && 2.3     && 1    \\
$\sigma_{z}(\mu$m$)$   & &72       && 44   \\\hline\hline
\end{tabular}
\end{table}

\begin{figure}[htbp!]
 \includegraphics[width=7cm]{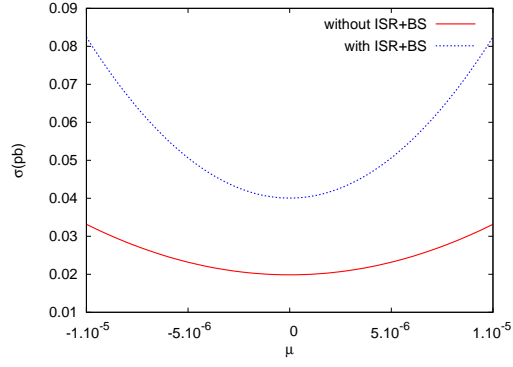}
 \caption{The total cross section of the process $e^-\gamma\to\nu\bar\nu e^-$ as a function of anomalous coupling
$\mu$ for the center-of-mass energy is taken to be $\sqrt s=3$ TeV
with and without ISR+BS effect.}\label{fig4}
\end{figure}

\begin{figure}[htbp!]
 \includegraphics[width=7cm]{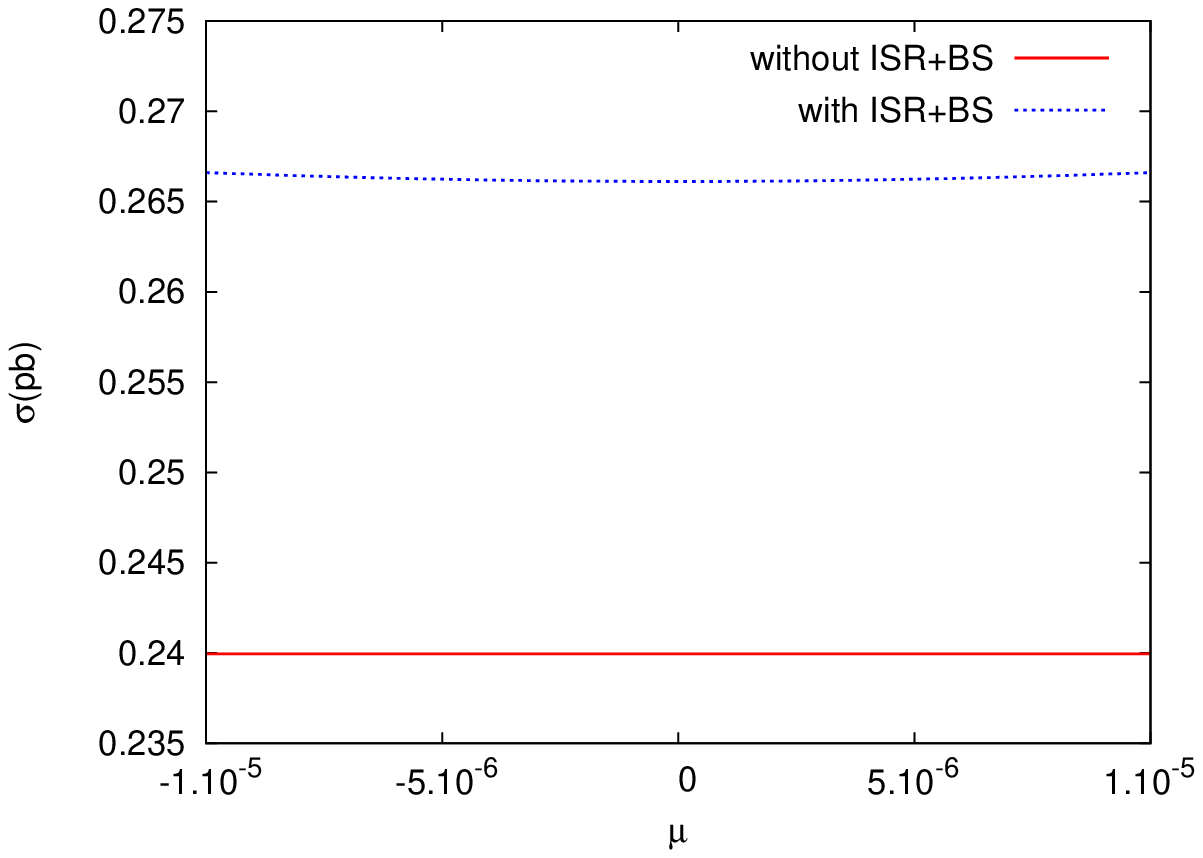}
 \caption{The same as Fig. \ref{fig4} but for $\sqrt s$=0.5 TeV.}\label{fig5}
\end{figure}

Now, we will analyze the $\nu\bar\nu\gamma$ coupling on the process
$e^-\gamma\to\nu\bar\nu e^-$ assuming neutrino magnetic moment
matrix is almost flavor diagonal ($\mu_{\nu_i}\gg\mu_{\nu_i\nu_j}$)
and only one of the matrix elements is different from zero
($\mu_{\nu_i}=\mu$). In Figs.\ref{fig4} and \ref{fig5}, we plot the
total cross sections for $e^-\gamma\to\nu\bar\nu e^-$  as a function
of anomalous coupling $\mu$ with and without ISR and beamstrahlung
effects for center of mass energies 3 TeV and 0.5 TeV, respectively.

\begin{figure}[htbp!]
 \includegraphics[width=7cm]{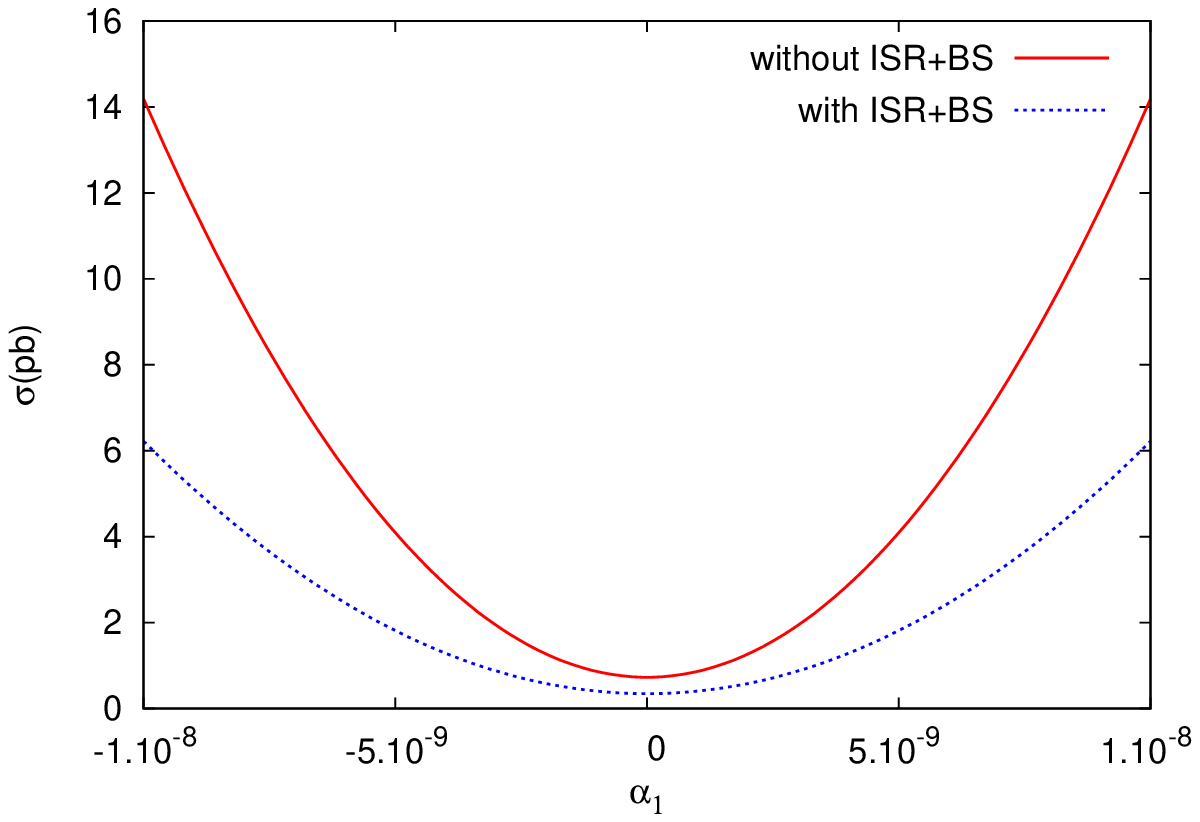}
 \caption{The total cross section of the process $e^-\gamma\to\nu\bar\nu e^-$ as a function of anomalous coupling
$\alpha_1$ for the center-of-mass energy is taken to be $\sqrt s=3$
TeV with and without ISR+BS effect.}\label{fig6}
\end{figure}

\begin{figure}[htbp!]
 \includegraphics[width=7cm]{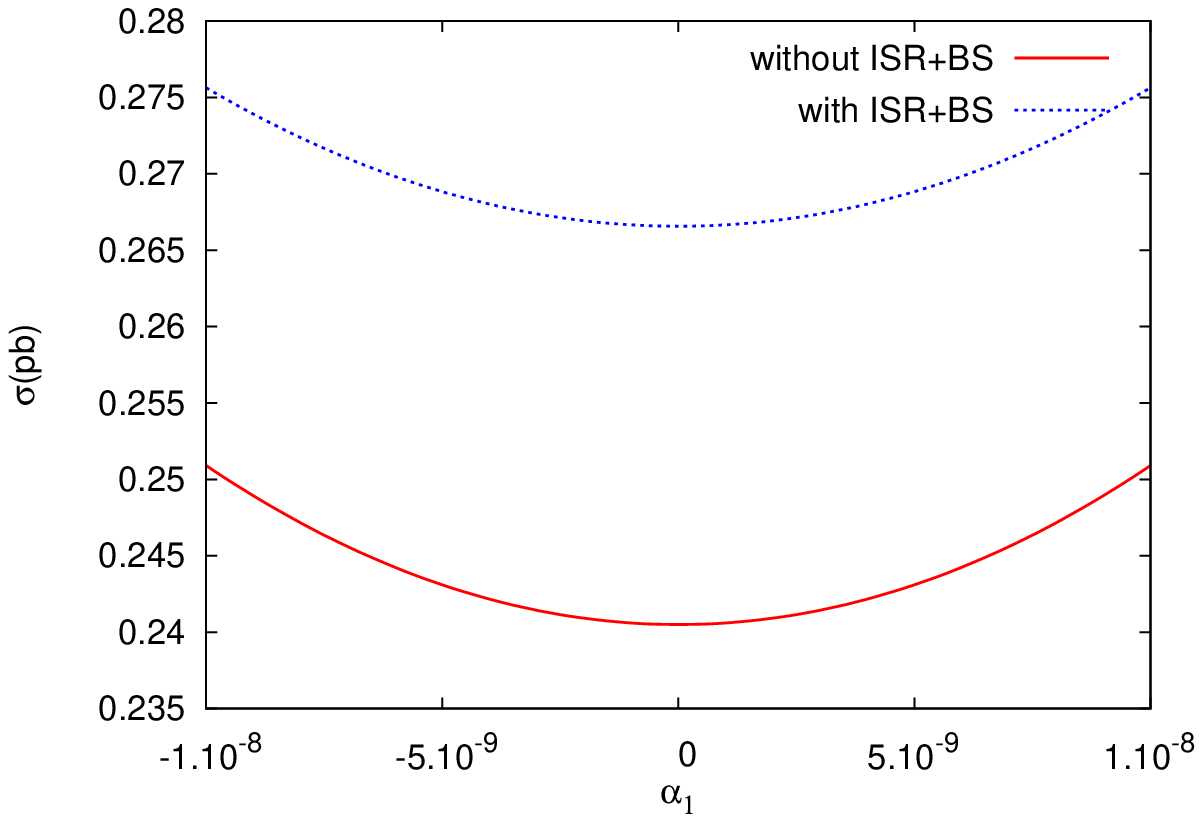}
 \caption{The same as Fig. \ref{fig6} but for $\sqrt s$=0.5 TeV.}\label{fig7}
\end{figure}

Fig.\ref{fig2} shows the total cross sections of
$e^-\gamma\to\nu\bar\nu e^-$  process depending on anomalous part as
function of $\alpha_1^2+\alpha_2^2$ are calculated in the case of
effective interaction (\ref{eqn2}). Here, $\alpha_1$ and $\alpha_2$
can written in the form:
\begin{eqnarray}
\label{alpha1}
\alpha_1^2=\sum_{i,j}\left[|\alpha^{ij}_{R1}|^2+|\alpha^{ij}_{L1}|^2\right]
,\;\;\;\;\;\;\;\alpha_2^2=\sum_{i,j}\left[|\alpha^{ij}_{R2}|^2+|\alpha^{ij}_{L2}|^2\right]
\end{eqnarray}
The total cross sections for $e^-\gamma\to\nu\bar\nu e^-$ as a
function of anomalous coupling $\alpha_1$ are plotted with and
without ISR and BS effects in Fig. \ref{fig6} at $\sqrt s$=3 TeV and
in Fig.\ref{fig7} at $\sqrt s$= 0.5 TeV.

\begin{figure}[htbp!]
 \includegraphics[width=14cm]{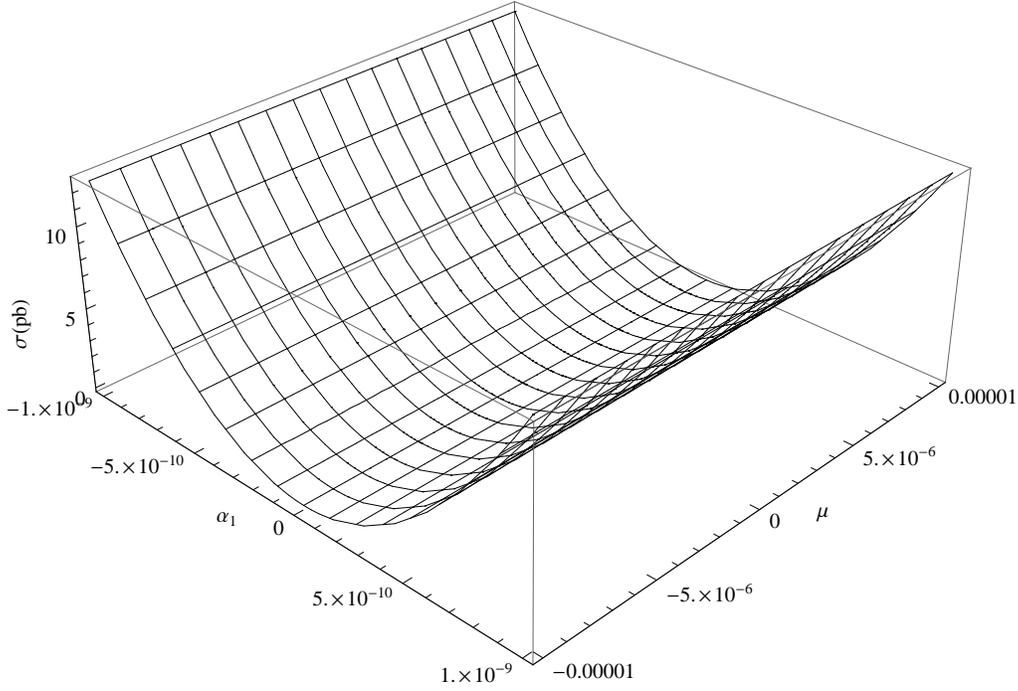}
 \caption{The total cross section of the process $\gamma\gamma\to\nu\bar\nu$ as a function of anomalous
 couplings $\mu$ and
$\alpha_1$ for the center-of-mass energy is taken to be $\sqrt s=3$
TeV.}\label{fig8}
\end{figure}

\begin{figure}[htbp!]
 \includegraphics[width=14cm]{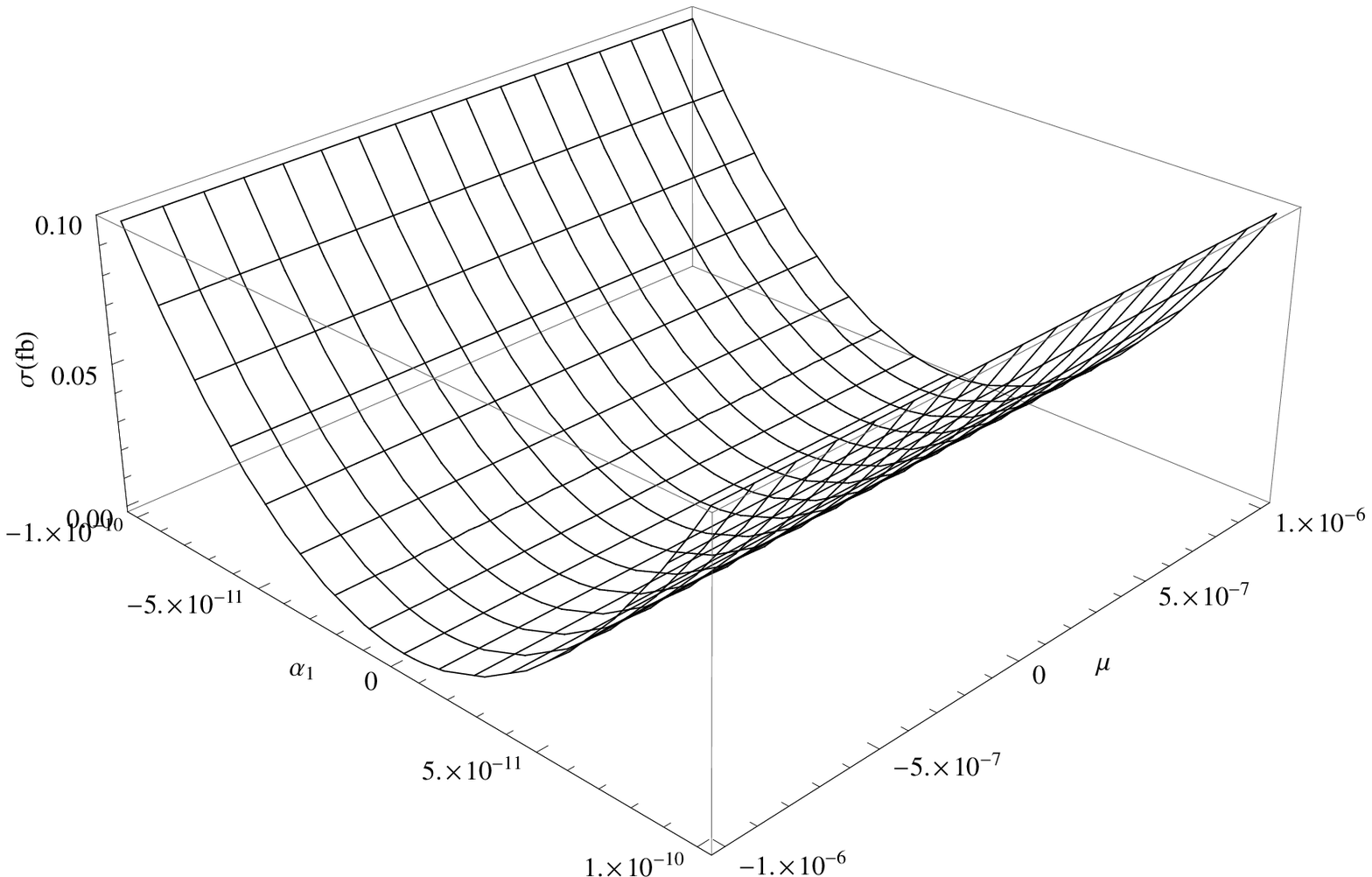}
 \caption{The same as Fig. \ref{fig8} but for $\sqrt s$=0.5 TeV.}\label{fig9}
\end{figure}

We have calculated analytical expression for the polarization summed
amplitude square for $\gamma\gamma\to\nu\bar\nu$ process with
CompHEP which agrees with reference \cite{Sahin:2010zr}. Although
the squared amplitude of the process depends on the anomalous
couplings both $\alpha_1$, $\alpha_2$ and $\mu$, we present the
total cross sections of $\gamma\gamma\to\nu\bar\nu$ as functions of
$\alpha_1$ and $\mu$ in Figs. \ref{fig8} for $\sqrt s$= 3 TeV and
\ref{fig9} for $\sqrt s$= 0.5 TeV at CLIC because of the fact that
the dependence of cross section on $\alpha_1$ and $\alpha_2$ is the
same.

\section{Numerical Results}
One-parameter $\chi^2$ test was applied without a systematic error
to obtain 95\% confidence level (C.L.) on the upper limits of the
$\alpha$ and $\mu$. The $\chi^2$ function is
\begin{eqnarray}
\chi^{2}=\left(\frac{\sigma_{SM}-\sigma_{AN}}{\sigma_{SM} \,\,
\delta}\right)^{2}
\end{eqnarray}
where, $\sigma_{AN}$ is the cross section containing new physics
effects and $\delta=\frac{1}{\sqrt{N}}$ is the statistical error.
The number of events are given by $N=\sigma_{SM}L_{int}$ where
$L_{int}$ is the integrated luminosity. Here, $\sigma_{SM}$ is the
corresponding SM background cross section. We have calculated
$\sigma_{SM}=3.73\times 10^{-2}$ pb at $\sqrt s=$ 3 TeV,
$\sigma_{SM}=2.66\times 10^{-1}$ pb at $\sqrt s=$ 0.5 TeV for
$e^-\gamma\to \nu_{\tau}\bar{\nu_{\tau}} e^-$ process.

As aforementioned, the neutrino magnetic moment matrix is almost
flavor diagonal, because the other elements in the matrix are
strictly constrained by the experiments. As well as, the element
$\mu_{\tau\tau}$ is dominant over the other diagonal elements
bounded with $3.9\times10^{-7} \mu_B$ \cite{Schwienhorst:2001sj}.
This bound is at least 3 orders of magnitude weaker than the bounds
on other diagonal matrix elements \cite{Davidson:2005cs}. Therefore,
we focus on $\mu_{\tau\tau}$ element of neutrino magnetic moment
matrix in our numerical calculations. In Tables \ref{tab1} and
\ref{tab2}, we present 95 \% C.L. upper bounds of the couplings
$\mu_{\tau\tau}$, $\alpha_1^2$ and $\alpha_2^2$ for two different
center of mass energies $\sqrt s$=3 TeV and $\sqrt s$=0.5 TeV,
respectively. When calculating the sensitivity of the anomalous
couplings, we take into account ISR+BS effect for incoming electron
and the backscattered photon distribution function for incoming
photon, and also a cut of $|\eta|< 2.5$ for pseudo-rapidity of final
electron for the process $e^-\gamma\to\nu\bar\nu e^-$. According to
these tables, our limits on $\mu_{\tau\tau}$ are the same order with
refs. \cite{Sahin:2010zr,Sahin:2012zm} but 5 times weaker than DONUT
bound \cite{Schwienhorst:2001sj}. Besides, our limits on
$\alpha_1^2$ and $\alpha_2^2$ ranged from order of $10^{-17}$ to
$10^{-21}$ at $\sqrt s$=0.5 and 3 TeV for $e^-\gamma\to\nu\bar\nu
e^-$ process as seen from Tables \ref{tab1} and \ref{tab2}. Our best
limits on $\alpha_1^2$ and $\alpha_2^2$ are 10 orders of magnitude
more restrictive than the LEP bound. On the other hand, it is 3
orders of magnitude more restrictive than LHC bounds
\cite{Sahin:2010zr,Sahin:2012zm}.

\begin{figure}[htbp!]
 \includegraphics[width=9cm]{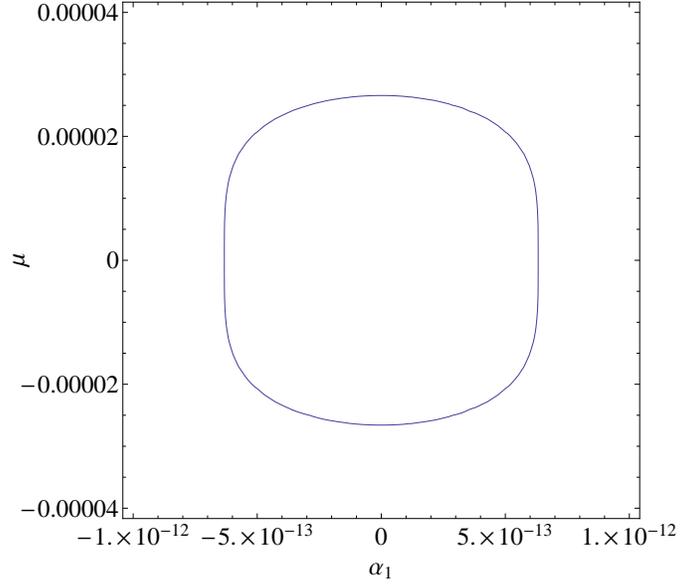}
 \caption{The contour plot for the upper bounds of the couplings $\mu_{\tau
 \tau}$ and
$\alpha_1^2$  with 95\% C.L. for the process
$\gamma\gamma\to\nu\bar\nu$ at $\sqrt s$= 3 TeV with corresponding
design luminosity.}\label{fig10}
\end{figure}

\begin{figure}[htbp!]
 \includegraphics[width=9cm]{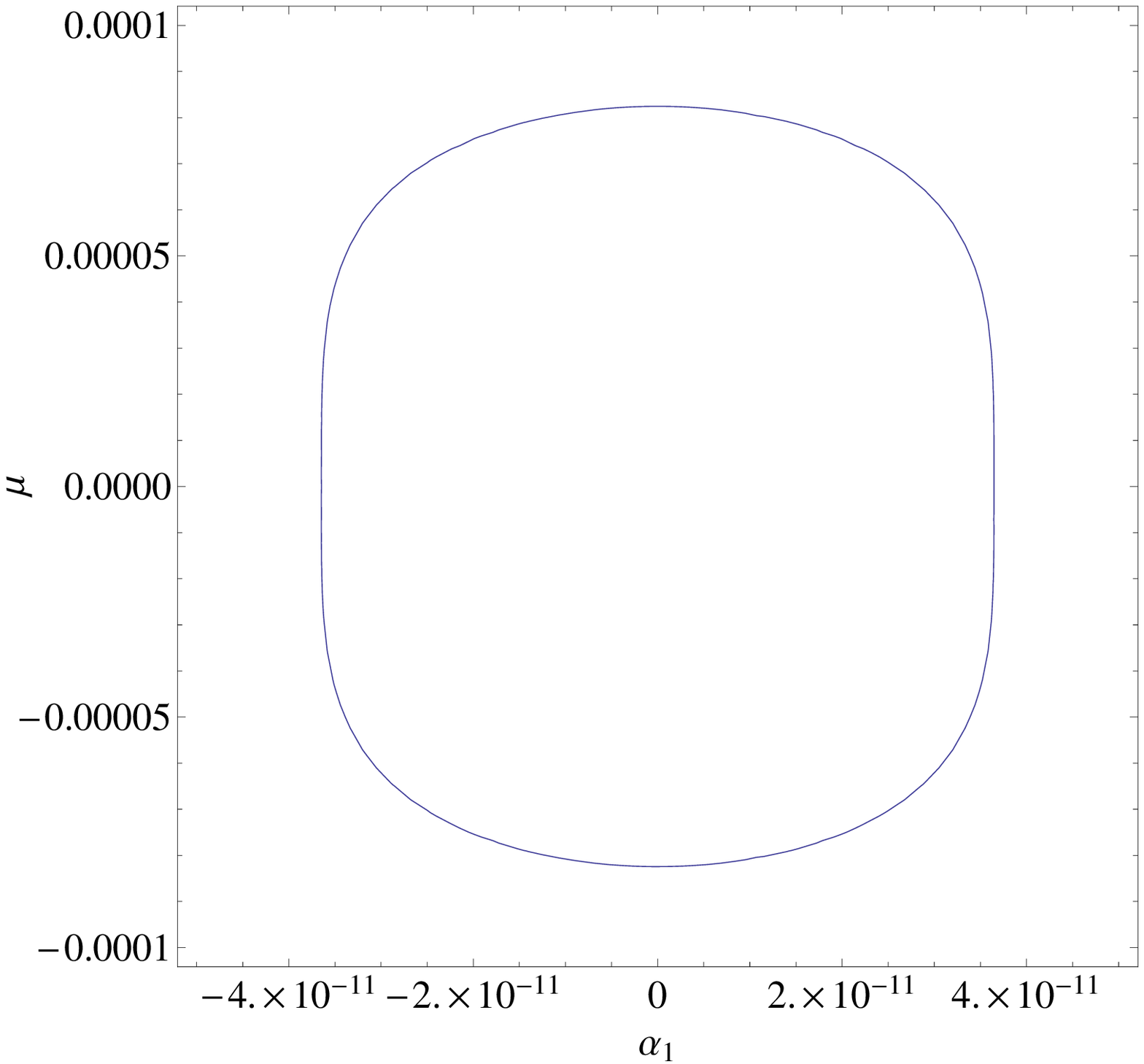}
 \caption{The same as Fig. \ref{fig10} but for $\sqrt s$=0.5 TeV.}\label{fig11}
\end{figure}

We used a Poisson distribution for searching sensitivity to
anomalous couplings through $\gamma\gamma\to\nu\bar\nu$ process.
Because, this process is absent in the SM at tree-level. In Figs.
\ref{fig10} and \ref{fig11}, we present the sensitivity contour plot
at \% 95 C.L. for the anomalous couplings, $\alpha_1$ and $\mu$,
through $\gamma\gamma\to\nu\bar\nu$ process with design luminosities
for $\sqrt s$= 3 and 0.5 TeV, respectively. In addition, we show 95
\% C.L. upper bounds of the couplings $\mu_{\tau\tau}$, $\alpha_1^2$
and $\alpha_2^2$ in Tables \ref{tab3} and \ref{tab4} for two
different center of mass energies $\sqrt s$=3 TeV and $\sqrt s$=0.5
TeV, respectively. As you can seen from these tables, our best
limits on $\alpha_1^2$ and $\alpha_2^2$ are about at order of
$10^{-25}$, but $\mu_{\tau\tau}$ limits are one order of magnitude
worse than the experimental current limits. As well as, our limits
on $\alpha_1^2$ and $\alpha_2^2$ can be reached at the order of
$10^{-21}$ at 10 $fb^{-1}$ luminosity with center of mass energy of
0.5 TeV while, in Ref. \cite{Sahin:2012zm} these limits are obtained
about at the order of $10^{-19}$ at LHC ($\sqrt s$=14 TeV) with 200
$fb^{-1}$ luminosity. And also, these limits are 15 orders of
magnitude more restrictive than LEP bound.
\section{Conclusions}
We have examined the potential of $e^-\gamma\to\nu\bar\nu e^-$ and
$\gamma\gamma\to\nu\bar\nu$ processes at the CLIC to search
neutrino-photon and neutrino-two photon couplings. The $e\gamma$ and
$\gamma\gamma$ collider modes of a linear collider probes
neutrino-photon and neutrino-two photon couplings with better
sensitivity than the present colliders. The neutrino-two-photon
vertex has important implications in astrophysics. It  contributes
to the process $\gamma \gamma \to \nu \bar \nu$ which is very
important in the energy-loss mechanism (cooling) in stars. It was
shown in Ref. \cite{Dodelson:1991dt} that stars, which have
intermediate temperatures and very low densities, are dominantly
affected by this process. We observe from Figs. \ref{fig8} and
\ref{fig9} that cross section of the process $\gamma \gamma \to \nu
\bar \nu$ is very sensitive to neutrino-two-photon coupling
$\alpha_1$ (as well as $\alpha_2$). Therefore, probing  the
couplings $\alpha_1$ and $\alpha_2$ is not only important for
understanding the physics beyond the Standard Model but also
contributes to the studies in astrophysics. We have shown that the
neutrino-two photon couplings improves the sensitivity limits by up
to a factor of $10^{15}$ with respect to LEP limits. Our limits are
also eight order of magnitude better than than the $\nu\bar \nu$
production in a $\gamma p$ collision at the LHC \cite{Sahin:2012zm}.
Besides, neutrino-photon coupling $\mu_{\tau\tau}$ are about an
order of magnitude worse than current experimental bound.
\begin{table}
\caption{95\% C.L. upper bounds of the couplings $\mu_{\tau \tau}$,
$\alpha_1^2$ and $\alpha_2^2$ for the process
$e^-\gamma\to\nu\bar\nu e^-$. We consider various values of the
integrated CLIC luminosities for $\sqrt s$=3 TeV. Limits of
$\mu_{\tau \tau}$ is given in units of Bohr magneton and $\Lambda$
is taken to be 1 GeV for limits of $\alpha_1^2$ and $\alpha_2^2$.
\label{tab1}}
\begin{ruledtabular}
\begin{tabular}{cccccc}
Luminosity:&10$fb^{-1}$& 30$fb^{-1}$ &50$fb^{-1}$ &100$fb^{-1}$&590$fb^{-1}$ \\
\hline
$\mu_{\tau \tau}$ &$2.68\times10^{-6}$ &2.04$\times10^{-6}$ &1.79$\times10^{-6}$ &1.51$\times10^{-6}$ &9.67$\times10^{-7}$ \\
$\alpha_1^2$&5.14$\times10^{-20}$ &2.97$\times10^{-20}$ &2.30$\times10^{-20}$ &1.63$\times10^{-20}$ &6.70$\times10^{-21}$   \\
$\alpha_2^2$&5.14$\times10^{-20}$ &2.97$\times10^{-20}$ &2.30$\times10^{-20}$ &1.63$\times10^{-20}$ &6.70$\times10^{-21}$  \\
\end{tabular}
\end{ruledtabular}
\end{table}
\begin{table}
\caption{The same as table \ref{tab1} but for $\sqrt s$=0.5 TeV.
\label{tab2}}
\begin{ruledtabular}
\begin{tabular}{cccccc}
Luminosity:&10$fb^{-1}$& 30$fb^{-1}$ &50$fb^{-1}$ &100$fb^{-1}$&230$fb^{-1}$ \\
\hline
$\mu_{\tau \tau}$ &$4.04\times10^{-5}$ &3.07$\times10^{-5}$ &2.70$\times10^{-5}$ &2.27$\times10^{-5}$ &1.84$\times10^{-5}$ \\
$\alpha_1^2$&8.89$\times10^{-17}$ &5.14$\times10^{-17}$ &3.98$\times10^{-17}$ &2.81$\times10^{-17}$ &1.85$\times10^{-17}$   \\
$\alpha_2^2$&8.89$\times10^{-17}$ &5.14$\times10^{-17}$ &3.98$\times10^{-17}$ &2.81$\times10^{-17}$ &1.85$\times10^{-17}$    \\
\end{tabular}
\end{ruledtabular}
\end{table}

\begin{table}
\caption{95\% C.L. upper bounds of the couplings $\mu_{\tau \tau}$,
$\alpha_1^2$ and $\alpha_2^2$ for the process
$\gamma\gamma\to\nu\bar\nu$. We consider various values of the
integrated CLIC luminosities for $\sqrt s$=3 TeV. Limits of
$\mu_{\tau \tau}$ is given in units of Bohr magneton and $\Lambda$
is taken to be 1 GeV for limits of $\alpha_1^2$ and $\alpha_2^2$.
\label{tab3}}
\begin{ruledtabular}
\begin{tabular}{cccccc}
Luminosity:&10$fb^{-1}$& 30$fb^{-1}$ &50$fb^{-1}$ &100$fb^{-1}$&590$fb^{-1}$ \\
\hline
$\mu_{\tau \tau}$ &$7.37\times10^{-5}$ &5.60$\times10^{-5}$ &4.93$\times10^{-5}$ &4.15$\times10^{-5}$ &2.67$\times10^{-5}$ \\
$\alpha_1^2$&2.36$\times10^{-23}$ &7.88$\times10^{-24}$ &4.73$\times10^{-24}$ &2.36$\times10^{-24}$ &4.00$\times10^{-25}$   \\
$\alpha_2^2$&2.36$\times10^{-23}$ &7.88$\times10^{-24}$ &4.73$\times10^{-24}$ &2.36$\times10^{-24}$ &4.00$\times10^{-25}$   \\
\end{tabular}
\end{ruledtabular}
\end{table}
\begin{table}
\caption{The same as table \ref{tab3} but for $\sqrt s$=0.5 TeV.
\label{tab4}}
\begin{ruledtabular}
\begin{tabular}{cccccc}
Luminosity:&10$fb^{-1}$& 30$fb^{-1}$ &50$fb^{-1}$ &100$fb^{-1}$&230$fb^{-1}$ \\
\hline
$\mu_{\tau \tau}$ &$1.81\times10^{-4}$ &1.37$\times10^{-4}$ &1.21$\times10^{-4}$ &1.02$\times10^{-4}$ &8.25$\times10^{-5}$ \\
$\alpha_1^2$&3.06$\times10^{-20}$ &1.02$\times10^{-20}$ &6.12$\times10^{-21}$ &3.06$\times10^{-21}$ &1.33$\times10^{-21}$   \\
$\alpha_2^2$&3.06$\times10^{-20}$ &1.02$\times10^{-20}$ &6.12$\times10^{-21}$ &3.06$\times10^{-21}$ &1.33$\times10^{-21}$    \\
\end{tabular}
\end{ruledtabular}
\end{table}

\begin{acknowledgments}
We would like to thank I.Sahin for useful comments and discussions.
We thanks G. Tavares-Velasco for valuable suggestions about the
implementation of $\nu\bar\nu\gamma\gamma$ vertex into the CompHEP
program.
\end{acknowledgments}

\end{document}